\documentclass[11pt,twoside]{article}

\usepackage{amsmath}
\usepackage{amssymb}
\usepackage{graphicx}
\usepackage{fancyhdr}
\usepackage{subfigure}

\fancypagestyle{plain}{%
\fancyhf{}%

}

\newcommand{\Cal}[1]{\ensuremath{\mathcal{#1}}}
\newcommand{\avg}[1]{\ensuremath{\langle{#1}\rangle_\Cal{D}}}

\newcommand{\uD}[1]{\ensuremath{#1_\Cal{D}}}
\newcommand{\uDnow}[1]{\ensuremath{#1_{\Cal{D}_0}}}
\newcommand{\Onow}{\Omega_{m0}}
\newcommand{\aD}{\ensuremath{a_\Cal{D}}}
\newcommand{\aDdot}{\ensuremath{\dot a_\Cal{D}}}
\newcommand{\aDddot}{\ensuremath{\ddot a_\Cal{D}}}
\newcommand{\dlz}{\ensuremath{D_{\rm L}(z)}}

\newcommand{\Sk}{\ensuremath{S_{k_\Cal{D}}}}
\newcommand{\DR}{\ensuremath{\Delta\Cal{R}}}

\begin{document}

\title{Explicit Cosmological Coarse Graining via Spatial Averaging}
\author{{\bf Aseem Paranjape}\footnote{E-mail:
    aseem@tifr.res.in}~~{\bf and}~{\bf T. P. Singh}
\footnote{E-mail: tpsingh@tifr.res.in},\newline\vspace{.1in}\\
{\it Tata Institute of Fundamental Research,}\\
{\it Homi Bhabha Road,}\\ {\it Mumbai 400005, INDIA.}}
\date{\today}

\maketitle

\thispagestyle{plain}

\begin{abstract}
\noindent The present matter density of the Universe, while highly
inhomogeneous on small scales, displays approximate homogeneity on
large scales. We propose that whereas it is justified to use the
Friedmann-Lema\^ itre-Robertson-Walker (FLRW) line element (which
describes an exactly homogeneous and isotropic universe) as a template
to construct luminosity distances in order to compare observations
with theory, the \emph{evolution} of the scale factor in such a
construction must be governed not by the standard Einstein equations
for the FLRW metric, but by the modified Friedmann equations derived by
Buchert \cite{buch1,buch4} in the context of spatial averaging in
Cosmology. Furthermore, we argue that this scale factor, defined in
the spatially averaged cosmology, will correspond to the effective 
FLRW metric \emph{provided} the size of the averaging domain coincides
with the scale at which cosmological homogeneity arises. This
allows us, in principle, to compare predictions of a spatially
averaged cosmology with observations, in the standard manner, for
instance by computing the luminosity distance versus red-shift 
relation. The predictions of the spatially averaged cosmology would in
general differ from standard FLRW cosmology, because the scale-factor
now obeys the modified FLRW equations. This could help determine, by
comparing with observations, whether or not cosmological
inhomogeneities are an alternative explanation for the observed cosmic
acceleration.   
\end{abstract}

\section{Introduction}
The goal of Cosmology is to describe the evolution history of the 
Universe as a whole, tracing it as far back as possible from the
present epoch. This seemingly daunting task is made tractable by the
fact that the observed matter density in the Universe appears to be
approximately homogeneous when viewed on the largest scales. Further,
observations of the Cosmic Microwave Background (CMB) radiation reveal
it to be isotropic (after subtracting a dipole contribution due to our
local motion) down to the level of one part in $10^5$. The twin
conditions of homogeneity and isotropy, if valid on all length scales,
would immensely simplify the form of the metric and matter tensors
used in the General Relativistic description of Cosmology. It can be
shown \cite{FRWref} that on purely geometric grounds, the
most general metric describing a homogeneous and isotropic universe
takes on the form of the Friedmann-Lema\^ itre-Robertson-Walker (FLRW)
line element. The standard method of proceeding is to assume that the
geometry of the universe can be described by a background metric with  
small perturbations, where the background is exactly homogeneous and 
isotropic and is hence of the FLRW form. (For a detailed discussion of
the assumptions involved in the standard Cosmology, see
Ref. \cite{ras}.)  

This approach is adopted not only for the early universe, for example
in describing Big Bang nucleosynthesis and the dynamics of the CMB
anisotropies, but is also assumed to be valid at recent epochs. In
particular, the luminosity distances constructed to compare data from
Type Ia Supernovae (SNe) with theory, assume a metric of the FLRW form
with its dynamics given by the Einstein equations for a homogeneous
and isotropic universe. We know, however, that the matter content of
the Universe today is \emph{not} homogeneous on small scales - one
finds inhomogeneity at the scale of individual galaxies for example;
and there is evidence \cite{voids} that voids of order $30h^{-1}$
Mpc\footnote{Here $h$ is the conventional parameter which appears in
  the definition of the Hubble constant $H_0=100h$
  kms$^{-1}$Mpc$^{-1}$.} in diameter account for $40$-$50\%$ of the
observed Universe. There is also evidence for voids 3-5 times this
size \cite{voids2}, as well as local voids on smaller scales
\cite{voids3}. The evidence for homogeneity of the matter distribution
in the Universe only appears at scales larger than about $100h^{-1}$
Mpc or so \cite{homog}. It is then reasonable to assume that the
metric describing our Universe must also reflect the inhomogeneity of
the matter on the corresponding scales. One may now ask whether it is
at all justified to use the luminosity distances constructed using the
FLRW metric with its associated dynamics to compare theory with
observational data from SNe. In this paper we argue that the use of
such luminosity distances is indeed justified, provided we are only
interested in using such constructions in probing length scales larger
than a certain minimum scale which is fixed by requiring that the
assumption of homogeneity be valid (a lower bound on such a scale
could be $100h^{-1}$ Mpc for example). More importantly, we propose
that the \emph{dynamics} of this (effective or template) macroscopic
metric must be governed not by the standard Einstein equations applied
to the FLRW metric, but by the effective equations obtained after
spatial averaging on constant time slices, as developed by Buchert
\cite{buch1,buch4}.    

An outstanding open issue has been the following: how does one compare 
the predictions of the inhomogeneous, spatially averaged cosmology
with observations, and should one be attempting an interpretation in
terms of spatially averaged variables in the first place? 
For instance, one could study light
propagation and determine the luminosity distance versus redshift
(\dlz) relation which best fits the observations, within an
\emph{inhomogeneous} model of the Universe.
Following the important and interesting early work of Dyer and Roeder,
where the angular-size and luminosity distances were obtained for
beams propagating in "lumpy" FLRW universes 
\cite{dyer-roeder},
detailed investigations have been carried out
in \cite{inhomdlz}.  Such a program has also
received considerable attention in the context of the spherically
symmetric inhomogeneous Lema\^itre-Tolman-Bondi (LTB) models
\cite{LTB,cel}. However, it is not clear to us that this is
necessarily the best way to proceed, at least as far as distant SNe
are concerned. A realistic computation of light
propagation in an inhomogeneous geometry seems difficult at best, and
while the use of simplified toy models such as the LTB solutions in this
endeavor may be justified on the grounds that we may inhabit a locally
underdense region \cite{voids2}, it is not clear that the assumptions
justifying the use of the toy model remain valid at large scales (see
also Ref. \cite{cel} for a recent review of the LTB approach). And
after all, the concordance model infers the presence of a dark energy
by comparing the predictions of the standard homogeneous and isotropic
FLRW cosmology against observations \cite{riess}. If one decides to
interpret the FLRW cosmology as a template against which observations
may be compared, then in order to decide whether cosmic
inhomogeneities could be an alternative to the concordance model, one
must construct a homogeneous and isotropic cosmology, after a suitable
smoothing procedure, and compare its predictions with observations
such as the \dlz\ relation for Type Ia SNe.     

One should note that the effects of averaging within the standard
perturbed FLRW framework of Cosmology, have been studied by several
authors (see, e.g., Refs. \cite{hirata,gruzinov}), and it has been
argued that these effects must necessarily be small. The papers in
Refs. \cite{hirata,gruzinov} however, do not necessarily address the
entire problem. For example, Ref. \cite{hirata} deals with only 
super-horizon fluctuations, whereas work done by others (see, e.g,
Ref. \cite{kolb}) indicates that there may be nontrivial contributions
due to averaging from \emph{sub}horizon fluctuations where the matter
perturbations have gone nonlinear. And while the authors of
Ref. \cite{gruzinov} construct a cosmological model with negligible
effects of averaging, there do exist toy models of inhomogeneous
spacetimes in which these effects are large (see, e.g.,
Ref. \cite{aptp}). 

Furthermore, the possibility of \emph{nonperturbative} effects of
averaging has not, to our knowledge, been satisfactorily addressed,
much less ruled out. It is important to realize that current $N$-body
simulations for example, which are based on a Newtonian approximation
of gravity, consistently miss possible relativistic effects of
averaging. This is because the modifications due to averaging in a
Newtonian setting can be shown to reduce to boundary terms in general
\cite{ehlers}, and to vanish in the case of spherical symmetry
\cite{kerscher}. These terms therefore do not contribute in
simulations which either set periodic boundary conditions on the
simulation box, or use a spherically symmetric simulation region. As
we will see, the ``backreaction'' effects in Buchert's framework
(i.e., the modifications to the Cosmological equations due to
averaging) are coupled to the average of the spatial curvature, this
being a completely relativistic effect. Nontrivial consequences of
this coupling deserve more attention than they have received in the
context of $N$-body simulations (see Ref. \cite{kerscher}).

We also note that an argument for describing the Universe using FLRW
plus Newtonian perturbations (e.g., as given in Ref. \cite{wald}), is
based on the fact that it is Newtonian perturbations \emph{and} a
negative pressure component, which describe the Universe very well. In
other words, one argues that the effect of averaging inhomogeneities
must be small, but at the cost of adding an ill-understood Dark Energy
component in the Universe. On the other hand one should allow for the
following possibility (as of now neither confirmed nor ruled out): 
at an initial epoch, say at the end of inflation, the perturbations in
the homogeneous metric would give a small ``backreaction'' as a result
of averaging. As a result of its coupling with the curvature, this
backreaction would grow and become significant around the time of
structure formation, when matter inhomogeneities on small scales
become nonlinear. The backreaction would also nontrivially affect the
growth of the metric perturbations. Such a possibility could only be
tested within a fully nonperturbative treatment of the evolution of
inhomogeneities.   

The goal of this paper is modest. We argue that the scale factor,
defined in the spatially averaged cosmology via the volume of a
domain, should correspond to an effective FLRW metric, provided the
size of the averaging domain coincides with the scale at which
cosmological homogeneity arises. This allows us, in principle, to
compare predictions of a spatially averaged cosmology with
observations, in the standard manner, for instance by computing the
\dlz\ relation. (However, see Ref. \cite{buchmorph} for an alternative 
approach to matching the spatially averaged dynamics with
observations.) The predictions of the spatially averaged cosmology 
would in general differ from standard FLRW cosmology, because the
scale-factor now obeys modified Einstein equations. We will
demonstrate that consistent models of spatially averaged FLRW
templates exist which can, at least in principle, explain observations 
such as those of the \dlz\ relation for Supernovae. 
\label{intro}

In a nutshell, our new results in this paper are that we propose a template 
metric which takes into account the kinematical effects of inhomogeneities in 
the Universe. To implement this idea, we first suggest that in Buchert's
spatial averaging scheme, the size of the averaging domain should coincide
with the cosmological scale at which homogeneity sets in. We then argue
that, as a first step towards predicting the relation between inhomogeneities
and observations, the expansion factor $a_{\cal D}(t)$ defined by Buchert
should be identified with the scale-factor $a(t)$ in standard FLRW cosmology.
The template metric is then the FLRW metric, with $a_{\cal D}(t)$ playing the
role of $a(t)$. We then show, with the help of a few toy examples, that the
luminosity distance - redshift relation in a model Universe with an 
accelerating $a_{\cal D}(t)$ can be computed in the standard manner. To the
best of our knowledge, such a template metric has not been proposed in
earlier work by other researchers.

\section{The Coarse Grained Picture and Cosmology}
At the length scales of cosmological interest, individual galaxies are 
small enough to be treated as point-like objects. Given the large
number ($\sim 10^{11}$) of galaxies estimated to exist in the
observed Universe, it is also assumed that the matter content of the
Universe can be described to a good approximation as a continuum
fluid. At the level of the matter distribution, this picture is
comparable to that of a gas composed of a large number of molecules,
wherein the internal structure of the molecules can be ignored when
dealing with the macroscopic properties of the gas which is treated as
a continuum fluid. Further, the complex intermolecular interactions in
the gas at the microscopic level can be dealt with in a mean field
approximation by considering volume averages of the quantities of
interest.  

This picture is made precise by introducing the concept of
\emph{physically infinitesimal} volumes \cite{lanfluid} or 
\emph{coarse-graining cells}. The fluid approximation is
assumed to be valid provided the (effectively) infinitesimal volumes
used in defining continuum quantities such as the matter density of
the fluid, correspond to \emph{physical} volumes which are large enough
to contain a very large number of the particles constituting the
fluid. At the same time, since these volumes are to be effectively
infinitesimal, their physical size must also be much smaller than the
scale of the fluid as a whole. We can then say that the length scales
$L$ being probed by these physically infinitesimal volumes must lie in
a certain range\footnote{In the cosmological context, these length
scales will be probed by the metric. Since we will soon specialize to
a metric written in comoving coordinates, these length scales must be
understood to be comoving scales.} $L_1<L<L_2$ which is determined by
the details of the fluid (see also \cite{buchell}). Clearly we must
satisfy $L_2\ll L_{syst}$ where $L_{syst}$ is the size of the system.
In the cosmological context, $L_{syst}$ would be the Hubble scale, and
in order to deal with homogeneous and isotropic effective descriptions
we would also have to satisfy $L_1\gtrsim L_{hom}$ where the
homogeneity scale $L_{hom}$ as mentioned earlier can be as large as
$100h^{-1}$ Mpc. The physical quantities we wish to deal with, such as
the matter density, pressure, fluid velocity field, etc. are defined
by suitably averaging or \emph{coarse graining} over the physically
infinitesimal volumes (hence the terminology of coarse-graining cells
(CGCs)). For example, the matter density of a given CGC is defined as
the mass contained in it divided by its physical volume. The
(effectively) continuum matter density is built up by repeating this
for every CGC in the fluid.  

\begin{figure}[t] 
\centering 
\includegraphics[width=.75\textwidth]{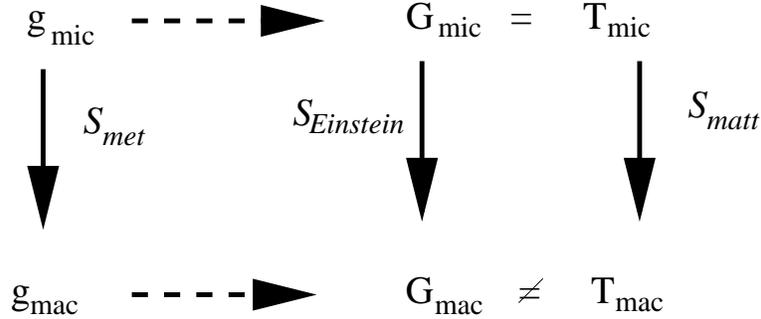}
\caption{Symbolic depiction of the modifications to Einstein's
  equations caused by an explicit smoothing. Adapted from Ellis
  \cite{ellis}, fig. 7. See text for details.}
\label{fig1}
\end{figure}

Since we deal with the notion of a continuum fluid in Cosmology, 
clearly we need a working description of coarse graining appropriate
in this context. However, the underlying theory being General 
Relativity, there are problems. The crucial issue is that along with 
the matter distribution (which is now tensorial), the \emph{metric}
for the spacetime also needs to be coarse grained. As discussed by
Ellis \cite{ellis}, in defining a smoothing operation to go from the
scale of stars (Scale 1 in Ellis' terminology, \emph{microscopic} in
ours) to cosmological scales where homogeneity can be assumed to have
set in (Scale 5 for Ellis, \emph{macroscopic} for us), one would
employ two logically different smoothing operators, one each on the
metric and the matter tensor. We denote the smoothing operator acting
on the metric as $S_{met}$ and the one acting on the matter tensor as
$S_{matt}$ (our notation differs from that used in
Ref. \cite{ellis}). We assume that Einstein's equations are valid on  
the microscopic scale of stars, which is reasonable since it is at 
these scales that Solar System tests and binary pulsar studies probe 
General Relativity. Symbolically then (see fig. \ref{fig1}), the  
metric $g_{mic}$ at this scale is used to construct the Einstein 
tensor $G_{mic}$ which is equated to the matter tensor $T_{mic}$, 
giving Einstein's equations $G_{mic}=T_{mic}$.  
To obtain the
corresponding description at the macroscopic scale, we use the
smoothing operators to get a new metric tensor $g_{mac} =
S_{met}(g_{mic})$ and a matter tensor $T_{mac}=S_{matt}(T_{mic})$. The
macroscopic metric $g_{mac}$ is used to construct the Einstein tensor
$G_{mac}$ at the macroscopic scale. Formally, one can think of the map
$S_{met}$ as inducing a map $S_{Einstein}$ which acts on the
microscopic Einstein tensor $G_{mic}$ to yield
$G_{mac}=S_{Einstein}(G_{mic})$. Since the operators $S_{Einstein}$
and $S_{matt}$ are in general different, we have $G_{mac}\neq
T_{mac}$. Hence, if the Einstein equations are valid on a given length 
scale, then they are in general not valid on some other length scale
connected to the first by appropriately defined smoothing operators.  
If we can re-express the new Einstein tensor $G_{mac}$ in terms of
quantities obtained by acting the operator $S_{matt}$ on \emph{both
  sides} of the Einstein equations at the microscopic scale, then it
should be possible to write $G_{mac}=T_{mac} + P_{mac}$ where
$P_{mac}$ represents corrections to the Einstein equations arising
from an explicit smoothing. In the standard approach, it is assumed
that these corrections are small enough to be neglected. (It should be
emphasized that given the recent evidence for voids in the Universe,
one or more additional scales may eventually have to be included in
the averaging scheme.)

Added to all this is the complication that smoothing or averaging
operations are notoriously difficult to define for tensor quantities
in a gauge covariant manner. Some examples of 
averaging techniques employed in General Relativity (and in Cosmology
in particular) can be found in Ref. \cite{avg}. Here we restrict
ourselves to the formalism developed by Buchert \cite{buch1,buch4} for
volume averages of scalar quantities in a given choice of
time-slicing. In the following sections we recall this formalism and
argue that it is naturally adapted to the notion of the smoothing
operator for the \emph{matter} sector. We will reserve the term
``averaging'' for the matter smoothing defined in this way. As for the
smoothing of the metric (which we will refer to as ``coarse
graining''), we will take recourse to cosmological observations and
argue that the result of such an operation on the inhomogeneous metric
must yield the FLRW metric on a suitable coarse graining scale. 
\label{cgcpic}

\section{Averaging in a Dust Dominated Spacetime}
For simplicity we restrict our attention to a pressure-less fluid or
``dust'', although a generalization to fluids with non-zero pressure
is also possible (see Ref. \cite{buch4}). For a general spacetime
containing irrotational dust, the metric can be written in synchronous
and comoving gauge\footnote{Latin indices take values 1..3,  Greek
  indices take values 0..3. We set $c=1$.},   
\begin{equation}
ds^2=\,-dt^2+h_{ij}(\vec{x},t)dx^idx^j\,.
\label{avg1}
\end{equation}
The expansion tensor $\Theta^i_j$ is given by $\Theta^i_j\equiv
(1/2)h^{ik}\dot h_{kj}$ where the dot refers to a derivative with
respect to (proper) time $t$. The traceless symmetric shear tensor is
defined as $\sigma^i_j\equiv \Theta^i_j-(\Theta/3) \delta^i_j$ where
$\Theta=\Theta^i_i$  is the expansion scalar. The Einstein equations
can be split \cite{buch1} into a set of scalar equations and a set of
vector and traceless tensor equations. The scalar equations are the
Hamiltonian constraint \eqref{avg2a} and the evolution equation for 
$\Theta$ \eqref{avg2b}, 
\begin{subequations}
\begin{equation}
^{(3)}\Cal{R}+\frac{2}{3}\Theta^2-2\sigma^2=16\pi G\rho\,,
\label{avg2a}
\end{equation}
\begin{equation}
^{(3)}\Cal{R}+\dot\Theta+\Theta^2=12\pi G\rho\,,
\label{avg2b}
\end{equation}
\label{avg2}
\end{subequations}
where $^{(3)}\Cal{R}$ is the Ricci scalar of the 3-dimensional
hypersurface of constant $t$, $\sigma^2$ is the rate of shear
defined by $\sigma^2\equiv(1/2)\sigma^i_j\sigma^j_i$ and $\rho$ is the
matter density of the dust. Note that all quantities in
Eqns. \eqref{avg2} generically depend on both position $\vec{x}$ and
time $t$. Eqns. \eqref{avg2a} and \eqref{avg2b} can be combined to
give Raychaudhuri's equation  
\begin{equation} 
\dot\Theta+\frac{1}{3}\Theta^2+2\sigma^2+4\pi G\rho=0\,. 
\label{avg3} 
\end{equation} 
The continuity equation $\dot\rho=-\Theta\rho$ which gives the 
evolution of $\rho$, is consistent with Eqns. \eqref{avg2a},  
\eqref{avg2b}. We only consider the scalar equations, since the
spatial average of a scalar quantity can be defined in a gauge 
covariant manner within a given foliation of space-time. For the 
space-time described by \eqref{avg1}, the spatial average of a scalar 
$\Psi(\vec{x},t)$ over a {\em comoving} domain \Cal{D} at time $t$ is 
defined by 
\begin{equation}
\avg{\Psi}=\frac{1}{\uD{V}}\int_\Cal{D}{d^3x\sqrt{h}\,\Psi}\,,
\label{avg4}
\end{equation}
where $h$ is the determinant of the 3-metric $h_{ij}$ and $\uD{V}$
is the volume of the comoving domain given by
$\uD{V}=\int_\Cal{D}{d^3x\sqrt{h}}$. Spatial averaging is, by
definition, not generally covariant. Thus the choice of foliation is
relevant, and should be motivated on physical grounds. In the context
of cosmology, averaging over freely-falling observers is a natural
choice, especially when one intends to compare the results with
standard FLRW cosmology.  Following the definition \eqref{avg4} the
following commutation relation then holds \cite{buch1}  
\begin{equation}
\avg{\Psi}^\cdot-\avg{\dot\Psi}=
\avg{\Psi\Theta}-\avg{\Psi}\avg{\Theta}\,,
\label{avg5}
\end{equation}
which yields for the expansion scalar $\Theta$
\begin{equation}
\avg{\Theta}^\cdot-\avg{\dot\Theta}=
\avg{\Theta^2}-\avg{\Theta}^2\,.
\label{avg6}
\end{equation}
Introducing the dimensionless scale factor
$\aD\equiv\left(\uD{V}/V_{\Cal{D} i}\right)^{1/3}$ normalized by the
volume of the domain \Cal{D}\ at some initial time $t_i$, we can
average the scalar Einstein equations \eqref{avg2a}, \eqref{avg2b} and
the continuity  equation to obtain \cite{buch1}
\begin{subequations}
\begin{equation}
\avg{\rho}^\cdot=\,-\avg{\Theta}\avg{\rho} ~~~;~~~
\avg{\Theta}=3\frac{\aDdot}{\aD}\,, 
\label{avg7a}
\end{equation}
\begin{equation}
3\left(\frac{\aDdot}{\aD}\right)^2-8\pi G\avg{\rho}+
\frac{1}{2}\avg{\Cal{R}}=\,-\frac{\uD{\Cal{Q}}}{2}\,,
\label{avg7b}
\end{equation}
\begin{equation}
3\left(\frac{\aDddot}{\aD}\right)+4\pi
G\avg{\rho}=\uD{\Cal{Q}}\,.
\label{avg7c}
\end{equation}
\label{avg7}
\end{subequations}
Here $\avg{\Cal{R}}$, the average of the spatial Ricci scalar 
$^{(3)}\Cal{R}$, is a domain dependent spatial constant. The
`kinematical backreaction' $\uD{\Cal{Q}}$ is given by  
\begin{equation}
\uD{\Cal{Q}}\equiv\frac{2}{3}\left(\avg{\Theta^2}-
\avg{\Theta}^2\right)-2\avg{\sigma^2}    
\label{avg8}
\end{equation}
and is also a spatial constant over the domain \Cal{D}.
%
%
%
A necessary condition for \eqref{avg7c} to integrate to
\eqref{avg7b} takes the form of the following differential equation
involving $\uD{\Cal{Q}}$ and $\avg{\Cal{R}}$,
\begin{equation}
\uD{\Cal{\dot Q}}+6\frac{\aDdot}{\aD}\uD{\Cal{Q}}
+\avg{\Cal{R}}^{\cdot}+2\frac{\aDdot}{\aD}\avg{\Cal{R}}=0\,.
\label{avg9}
\end{equation}
\label{avg}

\section{Smoothing Operators and Modified Field Equations}
The averager defined in Eqn. \eqref{avg4} is a natural candidate for 
the smoothing operator $S_{mat}$ which must act on the matter
tensor\footnote{We proceed with the understanding that we will only be
dealing with scalar quantities in a chosen time-slicing.}, although we
have not yet specified the domain size. Note that this
averaging is defined on an inhomogeneous underlying geometry,
i.e. it does not smoothen out the geometry itself. It remains to
determine the operator $S_{met}$ that will coarse grain the
metric. Although we do not have an explicit definition for this
operator, we can define a notional smoothing operator acting on the
metric, by using physical inputs from the Universe we observe. Namely,  
we propose that since the Universe has a matter content which is
homogeneous on large scales, and since the CMB is highly
isotropic\footnote{The CMB is isotropic \emph{after} subtracting a
  dipole contribution due to the local motion of our galaxy. This 
  motion can be thought of as a measure of inhomogeneities on small
  scales. The isotropy that we impose is therefore effectively an
  isotropy on large scales.} around us, this notional operator must
reflect these observations by acting on the general metric of
Eqn. \eqref{avg1} to yield a FLRW metric. Crucially, the \emph{scale
  factor} of this FLRW metric must (within our assumptions) be given by
Buchert's functional $\aD(t)$, where the domain size to be used will
be specified shortly. The argument leading to this proposal is as
follows.   

We begin by making precise the notion of ``homogeneity
on large scales'', by requiring that in the range $R\equiv(L_1,L_2)$
of CGC sizes allowed by the fluid approximation validity criterion,
there exist a subset $R_{hom}\equiv\subset R$ satisfying the following
condition: When the microscopic matter density $\rho(\vec{x},t)$ is
averaged over a length scale $L\in R_{hom}$, the resulting smoothed
density \emph{must lose all information} about the particular CGC over
which the smoothing was performed. In other words, for $L\in R_{hom}$,
at any given time $t$, the smoothed matter density must have the
\emph{same} value in all CGCs. It is in this sense that the smoothed
matter density is homogeneous (see however, the discussion at the end
of this section concerning an explicit smoothing of the metric). This
definition can be related to the concept of \emph{statistical}
homogeneity if we assume that the total volume of the observable
Universe is large enough that each CGC effectively samples one of an
ensemble of configurations of matter distribution (see
Ref. \cite{stoeg} for a careful discussion of this point). It is
possible that homogeneity sets in at a scale too large to be
considered as the size of the CGCs, namely that the range $R_{hom}$ is
not contained in the range $R$ and that $L_2^\prime\ll L_{\rm Hubble}$
is not satisfied. We will assume that this does not happen (for a more
detailed discussion on this assumption, see Sec. 2.1 of
Ref. \cite{ras}). We also make the crucial assumption that $R_{hom}$
does not change with time. This will later allow us to consistently
analyze light propagation using the effective FLRW metric. This
assumption needs some justification and we will return to this issue
in the discussion.   

Now define a coarse-graining cell to be an averaging domain \Cal{D}\
as used in Buchert's formalism, with a characteristic comoving size
$L\in R_{hom}$. The averaged density is then  $\avg{\rho}(t)$,
independent of the location of the particular CGC being considered. It
also follows that the averaged expansion scalar
$\avg{\Theta}=3(\aDdot/\aD)=( \uD{\dot V}/\uD{V})$ is homogeneous in
the above sense (using Eqn. \eqref{avg7a}) and describes the evolution
of the physical volume of any given CGC. Thus, $\avg{\Theta}$ plays
the role of the expansion scalar for a geometry whose infinitesimal
spatial volume elements probe length scales of the size of our CGCs
(see however the discussion at the end of this section). It is
therefore natural to demand that $\avg{\Theta}$ be the expansion
scalar for our notionally smoothed out macroscopic metric. This is
equivalent to demanding that the scale factor of this homogeneous and
isotropic effective metric be identified with the functional $\aD(t)$
(up to a constant factor which we choose to be unity by an appropriate
choice of units for the effective comoving coordinates). We are thus
left with an effective metric whose line element can be written as  
\begin{equation}
ds^2_{\rm eff}=-dt^2+\aD(t)^2\left(\frac{dr^2}{1-\uD{k} r^2} +
r^2d\Omega^2\right)\,,
\label{smop1}
\end{equation}
which probes comoving length scales of size $L$ as chosen above.
The 3-Ricci scalar of this effective metric is given by $\Cal{R}_{\rm
  eff} = 6\uD{k}/\aD^2$ and should be thought of as arising from the
coarse graining of the geometry. Following Ellis' ideas from
Sec. \ref{cgcpic} then, $\Cal{R}_{\rm eff} (=
S_{Einstein}(^{(3)}\Cal{R}))$ is in general different 
from the averaged 3-Ricci scalar $\avg{\Cal{R}}$ which arises from
applying the matter averager to the inhomogeneous scalar
curvature ($\avg{\Cal{R}}=S_{matt}(^{(3)}\Cal{R})$). The constant
$\uD{k}$ has the usual interpretation as a parameter characterizing
all possible  homogeneous and isotropic 3-metrics. We can write
\avg{\Cal{R}}\ in terms of $\Cal{R}_{\rm eff}$ by defining
$\DR\equiv\avg{\Cal{R}}-6\uD{k}/\aD^2$. In doing so we are merely
shifting focus from (the \emph{a priori} unknown) \avg{\Cal{R}}\ to
another unknown {\DR}. As the notation suggests, we think of \DR\ as
being a correction (a part of $P_{mac}$ in the notation of
Sec. \ref{cgcpic}, see Eqns. \eqref{smop2} below) to the 3-Ricci
scalar of the FLRW metric, although this correction need not be small
compared to $\Cal{R}_{\rm eff}$ and need not evolve proportional to
$\aD^{-2}$. We choose not to normalize the magnitude of the constant
\uD{k}, retaining instead the normalization of the functional $\aD(t)$
at the initial time $t_i$. The evolution equations appropriate at this
scale are Buchert's equations \eqref{avg7b} and \eqref{avg7c} together
with the integrability condition \eqref{avg9}. For completeness, and
in the spirit of Ellis' arguments we can rewrite equations
\eqref{avg7b} and \eqref{avg7c} as follows     
\begin{subequations}
\begin{align}
\underbrace{\left(\frac{\aDdot}{\aD}\right)^2+\frac{\uD{k}}{\aD^2}}~~&=~~ 
\underbrace{\frac{8\pi G}{3}\avg{\rho}}~~ - ~~
\underbrace{\frac{1}{6}\left(\DR +\uD{\Cal{Q}}\right)}
\label{smop2a}\\
G_{mac}~~~~~~~&~~~~~~~~~~~T_{mac}
~~~~~~~~~~~~~~~~~P_{mac}\nonumber\\ 
\overbrace{~~~~\frac{\aDddot}{\aD}~~~~}~~~&=~\overbrace{-\frac{4\pi
  G}{3}\avg{\rho}}~+~\overbrace{~~~~~~\frac{1}{3}
  \uD{\Cal{Q}}~~~~~~} \label{smop2b}   
\end{align}
\label{smop2}
\end{subequations}
where we have symbolically denoted the specific components of the
general modified equations to which the various terms of Buchert's
equations belong. The integrability condition \eqref{avg9} which
supplements these equations is purely a consequence of having non-zero 
corrections $P_{mac}$ to the standard FLRW equations. As a quick check, 
we note that requiring the corrections $P_{mac}$ to vanish implies
both $\uD{\Cal{Q}}=0$ (from Eqn. \eqref{smop2b}) and $\DR=0$, i.e. 
$\avg{\Cal{R}}=6\uD{k}/\aD^2$ (from Eqn. \eqref{smop2a}); the
integrability condition \eqref{avg9} is identically satisfied and we
recover the usual FLRW solution.

In the arguments leading to Eqns. \eqref{smop2}, we have ignored the
effects of a virtual \emph{change} in the averaging length scale
$L$. Considerable effort has been spent by Buchert and Carfora
\cite{buchRG} in studying such effects. In particular, in the spirit
of the real space Renormalization Group formalism, these authors
derive a novel \emph{curvature backreaction} effect analogous to the
kinematical backreaction \uD{\Cal{Q}}, together with a volume scaling
effect. The inclusion of these effects would change the evolution
equations satisfied by the scale factor \aD. We also note that Buchert
and Carfora's techniques (involving Ricci flow of the inhomogeneous
geometry into a smooth geometry) form one candidate for an explicit
realization of the metric coarse graining operator $S_{met}$. Seen in
this light, it is apparent that by accounting for the above effects,
we would also have to appropriately modify our notion of large scale
homogeneity (to include the volume scaling effect and the residual
information from the curvature fluctuations). In the arguments leading
to the effective metric \eqref{smop1}, the scale factor \aD\ would be 
replaced by an appropriately ``dressed'' quantity characterizing the
evolution of effectively infinitesimal volumes in the smoothed
geometry. However, the main arguments leading to an effective metric
which is FLRW, will not be altered given the existence of the range
$R_{hom}\subset R$. For simplicity, we ignore these additional effects
in this paper, and hope to investigate them in future work.     

It should be noted that a notion of coarse-graining is always implicit
in standard FLRW cosmology. It is assumed that a coarse-graining (while
satisfying the conditions of the fluid approximation) is possible on a
length scale which coincides with the scale at which homogeneity is
achieved (in other words, the existence of $R_{hom}\subset R$ is
assumed). Standard cosmology however ignores the possible
modifications to FLRW cosmology brought about by the process of
explicitly smoothing over the inhomogeneities - an issue which
Buchert's approach addresses. Our key proposal in this paper is that
the size of the averaging domain should be taken to coincide with the
scale of homogeneity, and that (up to some corrections) the quantity
$\aD(t)$ corresponds to the scale factor of a template FLRW metric.   

We now illustrate how our proposal can facilitate the comparison of
the predictions of the spatially averaged cosmology with observations,
for example by computing the \dlz\ relation using the scale factor
$\aD(t)$, which has a modified time evolution compared to the
corresponding scale factor in standard cosmology.  
\label{smop}

\section{Luminosity Distance using the Effective Metric}
In our approach, we take the effective or template metric to be FLRW,
and since we are holding the size $L$ of the averaging domain fixed,
there is nothing conceptually new to be done while studying its null
geodesics. The important difference, of course, is that the evolution
of the scale factor is now given by the modified Friedmann equations
\eqref{smop2} (subject to the caveats listed at the end of the
previous section). We note that the use of the FLRW metric is,
strictly speaking ``unrealistic'' in the sense that in the real
Universe, light travels mostly in vacuum which has a non-zero Weyl
tensor whereas in the FLRW cosmologies which are conformally flat,
light travels through a geometry with vanishing Weyl tensor (see also
Refs. \cite{dyer-roeder} and \cite{ras}). This issue is clearly relevant 
in the standard
approach as well. While we have not addressed this issue in the
context of averaging, we emphasize that our approach must be thought
of as a first step in gauging the effects of inhomogeneities, by using
an \emph{explicitly} averaged construction of the template FLRW model
to be used. The arguments relating the redshift of a source to the 
scale factor \aD\ and to its comoving distance $r$ from the observer,
are exactly the same as in the standard approach. In particular, we
have for the redshift $z$ of a source that emits light at time $t$
which is received by an observer at time\footnote{The subscript $0$
  will indicate the value of the quantity at the present epoch $t_0$,
  and the subscript $i$, the value at the initial epoch $t_i$.} $t_0$,    
\begin{equation}
1+z=\frac{a_{\Cal{D}_0}}{\aD(t)}\,,
\label{lumin1}
\end{equation}
and the comoving distance $r$ to the source is determined by solving
the equation
\begin{equation}
\int_0^r{\frac{dr}{\sqrt{1-\uD{k} r^2}}}\,=\,\int_t^{t_0}
    {\frac{dt^\prime}{\aD(t^\prime)}}\,.
\label{lumin2}
\end{equation}
A caveat to the above relations is that the ``step-size'' used
in computing the integrals involved must be understood to be no
smaller than the size $L$ of the CGCs. Further, the entire
construction must be interpreted as a ``fitting template'' that allows
us to compare observations with theoretical predictions in a straight
forward manner. With our choice of conventions, the magnitude of
\uD{k} is a dynamically determined parameter of the model and
determines the form of the luminosity distance \dlz\ as
follows. Solving Eqn. \eqref{lumin2} for $r=r_{em}(z)$ gives   
\begin{align}
r_{em}(z)&=\Sk\left(\alpha\left(z\right)\right) ~~;~~
\alpha(z)=\int_{\aD(z)}^{\uDnow{a}}{\frac{d\aD}{\aD^2\uD{H}}}\,,
\nonumber\\ 
\dlz&=\uDnow{a}(1+z)r_{em}(z)\,,
\label{lumin3}
\end{align}
where $\uD{H}=\aDdot/\aD$ is the Hubble parameter and the function \Sk\
is defined as
\begin{equation}
\Sk(x)=\left\{
\begin{array}{l}
(1/\sqrt{\uD{k}})\sin\left(\sqrt{\uD{k}}x\right) ~,~~~~~~~~\uD{k}>0\\
x~,~~~~~~~~\,~~~~~~~~~~~~~~~~~~~~~~~~~\uD{k}=0\\
(1/\sqrt{-\uD{k}})\sinh\left(\sqrt{-\uD{k}}x\right) ~,~~\uD{k}<0\, .
\end{array}\right .
\label{lumin4}
\end{equation}
In order to complete the picture we need the functional dependence of
the Hubble parameter \uD{H}\ on the scale factor \aD\ . This requirement
is complicated by the fact that the system of equations \eqref{avg7a},
\eqref{avg7b}, \eqref{avg7c} and \eqref{avg9} is only consistent, it
does not close. In order to obtain the required relation then, it is
necessary to make certain assumptions about the evolution of the
kinematical backreaction $\uD{\Cal{Q}}$ and the averaged 3-Ricci
scalar \avg{\Cal{R}}. Buchert \emph{et al.} \cite{buchmorph} have 
analyzed a class of scaling solutions of the form
\begin{equation}
\Cal{Q}_{\Cal{D}}=\Cal{Q}_{\Cal{D}_i}\aD^n ~~~;~~~
\avg{R}=\Cal{R}_{\Cal{D}_i}\aD^p\,.
\label{lumin5} 
\end{equation}
The case $n=-6$, $p=-2$ is the only scaling solution with $n\neq
p$. For the cases $n=p$, the backreaction is proportional to the
averaged 3-curvature\footnote{We have retained the notation of
  Ref. \cite{buchmorph}. The proportionality constant $r$ should not be
  confused with the comoving coordinate in the effective metric.},
\begin{equation}
\Cal{Q}_{\Cal{D}_i}=r\Cal{R}_{\Cal{D}_i} ~~~;~~~ r=-\frac{n+2}{n+6}
~~~;~~~ n=-2\frac{1+3r}{1+r}\,.  
\label{lumin6}
\end{equation}
The forms of the integrability condition \eqref{avg9} and the
solutions \eqref{lumin5} indicate that the scaling solutions can be
superposed to obtain new solutions. We will use two such
superpositions to construct models which contain an accelerating scale
factor \aD\ and which also yield analytic expressions for the
luminosity distance {\dlz}. The solutions we consider are of the form 
\begin{equation}
\avg{\Cal{R}}= \frac{6\uD{k}}{\aD^2} + \frac{\beta}{r}\aD^n ~~;~~ 
\uD{\Cal{Q}}=\beta\aD^n\,,
\label{lumin7}
\end{equation}
in keeping with our earlier discussion of the quantity {\DR}. Namely,
we model \DR\ by a single scaling solution. Here $r$ and $n$ are
related as in Eqn. \eqref{lumin6}, and \uD{k} and $\beta$ are
constants with dimensions of (length)$^{-2}$. We are interested in
models which admit an accelerating scale factor {\aD}, and will hence
assume $\beta>0$ and $\uD{k}<0$ (the reason for which will be apparent
shortly). Inserting the relations \eqref{lumin7} in
Eqns. \eqref{smop2a} and \eqref{smop2b} we find 
\begin{subequations}
\begin{equation}
\frac{\aDdot^2}{\aD^2}+\frac{\uD{k}}{\aD^2} = \frac{8\pi
  G}{3}\frac{\rho_i}{\aD^3}+\frac{2}{3}\frac{\beta}{n+2}\aD^n\,,
\label{lumin8a}
\end{equation}
\begin{equation}
\frac{\aDddot}{\aD}=-\frac{4\pi
  G}{3}\frac{\rho_i}{\aD^3}+\frac{\beta}{3}\aD^n\,,
\label{lumin8b}
\end{equation}
\end{subequations}
where we have defined $\rho_i\equiv\avg{\rho}(t_{in})$ and hence
written $\avg{\rho}=\rho_i/\aD^3$ using the continuity equation
\eqref{avg7a}. The next two examples which we will consider are
pathological, in that the luminosity distances \dlz\ constructed for
these models are not defined for all $z>0$. Nevertheless, since these
models yield analytical results, we shall display the expressions for
\dlz\ that we obtain. In general the computation of \dlz\ expressions
would have to be performed numerically.\\~\\

\noindent{\it Case 1} : $n=-3$. The backreaction $\uD{\Cal{Q}}$ in  
this model decays at the same rate as the averaged matter density 
$\avg{\rho}$. We choose initial conditions such that $\beta>4\pi
G\rho_i$, which ensures that the model accelerates 
(cf. Eqn. \eqref{lumin8b}). This is consistent with
Eqn. \eqref{lumin8a} due to the presence of the term 
$\uD{k}\aD^{-2}<0$, which allows the right hand side to be
negative. Defining $A\equiv (2/3)(\beta-4\pi G\rho_i)>0$ we find  
\begin{equation}
\aD^2\uD{H}= \aD\aDdot=\sqrt{\aD}\left[(-\uD{k})\aD-A\right]^{1/2}\,. 
\label{lumin9}
\end{equation}
Defining the ``volume deceleration''
$\uD{q}\equiv-\aDddot/(\aD\uD{H}^2)$ and evaluating
Eqns. \eqref{lumin8a} and \eqref{lumin8b} at the present epoch $t_0$,
we find  
\begin{align}
\uDnow{q}= \frac{\Onow}{2}-\frac{\uDnow{\Cal{Q}}}{3\uDnow{H}^2}&=
-\frac{1}{2}\frac{A}{\uDnow{a}^3\uDnow{H}^2}<0\,;\nonumber\\ 
\uD{k}= \left(\uDnow{a}\uDnow{H}\right)^2\left(\Onow-1-
\frac{2\uDnow{\Cal{Q}}}{3\uDnow{H}^2} \right)&=
-\left(\uDnow{a}\uDnow{H}\right)^2\left(1-2\uDnow{q}\right)<0\,,
\label{lumin10}
\end{align} 
where $\uDnow{\Cal{Q}}=\beta/\uDnow{a}^3$ is the present value of the
backreaction, and we have defined the present matter density parameter
$\Onow=(8\pi G\rho_0)/(3\uDnow{H}^2)=(8\pi
G\rho_i)/(3\uDnow{a}^3\uDnow{H}^2)$ in the usual way. The function
$\alpha(z)$ defined in Eqn. \eqref{lumin3} which determines the
luminosity distance \dlz\ reduces to 
\begin{equation}
\alpha(z)=\frac{2}{\sqrt{-\uD{k}}}
\ln\left[\frac{\left(1+\sqrt{1-2\uDnow{q}}\right)\sqrt{1+z}}
  {\sqrt{1-2\uDnow{q}}+\sqrt{1+2\uDnow{q}z}}\right]\,.  
\label{lumin11}
\end{equation}
Since $\uDnow{q}<0$, the redshift is constrained to take values $0\leq
z\leq 1/(-2\uDnow{q})$. \\~\\

\noindent{\it Case 2} : $n=-5/2$. Here the backreaction \uD{\Cal{Q}}\
decays slower than the matter density, and if we assume $0<\beta<4\pi
G\rho_i$, then we have a situation wherein the effective scale factor
initially decelerates, and then starts accelerating after a certain
epoch which is determined by the values of the parameters $\beta$ and
$\rho_i$. Using Eqn. \eqref{lumin8a} we have
\begin{equation}
\aD\aDdot=\sqrt{\aD}\left(\frac{8\pi G}{3}\rho_i-\frac{4\beta}{3}\sqrt{\aD}
-\uD{k}\aD\right)^{1/2}\,.   
\label{lumin12}
\end{equation}
Now we have $\uDnow{q}=(\Onow/2)-\uDnow{\Cal{Q}}/(3\uDnow{H}^2)$ as
before with $\uDnow{\Cal{Q}}=\beta/\uDnow{a}^{5/2}$, and
\begin{equation}
-\uD{k}=\left(\uDnow{a}\uDnow{H}\right)^2\left(1
 +\Onow-4\uDnow{q}\right)\,. 
\label{lumin13}
\end{equation}
The volume deceleration \uDnow{q} will be negative provided
\uDnow{a} corresponds to a sufficiently late epoch, namely if
$\uDnow{a}>(4\pi G\rho_i/3\beta)^2$ as can be seen from
Eqn. \eqref{lumin8b}, and provided $\uD{k}<0$ as seen from 
Eqn. \eqref{lumin13}. Assuming these conditions are met, we find for   
$\alpha(z)$ 
\begin{align}
\alpha(z)&=\frac{2}{\sqrt{-\uD{k}}}
\bigg\{\ln\left[\left(C+1-2\uDnow{q}\right)\sqrt{1+z}\,\right]
\nonumber\\ 
&~~~-\ln\left[C^2
    -\left(\Onow-2\uDnow{q}\right)\sqrt{1+z}
    +C\left\{1-2\left(\Onow-2\uDnow{q}\right)\left(
    \sqrt{1+z}-1\right)+\Onow z\right\}^{1/2}\right]\bigg\} \nonumber\\ 
C&\equiv\sqrt{1+\Onow-4\uDnow{q}}\,.
\label{lumin14}
\end{align} 
The range of allowed values for $z$ is constrained in this case as
well. For example, the quantity within the smaller braces in the
previous equation must be positive, and this can be shown to imply
that 
\begin{equation}
(z-z_+)(z-z_-)>0\,,
\label{lumin15}
\end{equation}
where $z_\pm$ are the roots of the quadratic polynomial
\begin{equation}
\Cal{P}(z)\equiv M_2^2z^2+(2M_1M_2-1)z+M_1^2-1 ~~;~~
M_1=1+\frac{1}{2\left(\Onow-2\uDnow{q}\right)} ~~;~~
M_2=\frac{\Onow}{2\left(\Onow-2\uDnow{q}\right)}\,.
\label{lumin16}
\end{equation}
An analysis of this condition reveals that unless
$\Onow>4\uDnow{q}^2$, the range $0<z_-<z<z_+$ must be excluded from
consideration when $\uDnow{q}<0$. Requiring the arguments of the
logarithms in Eqn. \eqref{lumin14} to be positive would also impose
some condition on the allowed values for $z$. In any case, since there
is no \emph{a priori} reason to expect that $\Onow>4\uDnow{q}^2$
holds, we see that this model also contains a pathology in
general.

Although both models considered above must be regarded as unphysical
(since the luminosity distance of a source is a measurable quantity),
consistent models also exist. For example, the case $n=0$, $r=-1/3$
corresponding to a constant backreaction, mimics the cosmological
constant of the standard cosmology. Further, it may be argued that
current observations seem to indicate that we live in a universe with
$\uD{k}=0$ and hence the above examples are not observationally
relevant in any case. To show that the formalism does admit consistent
(accelerating) models with $\uD{k}=0$ as well, we now give an example
of such a model. \\~\\

\noindent{\it Case 3} : $n=-1$, $\uD{k}=0$. In this case, although the
\emph{effective} spatial curvature $\Cal{R}_{eff}$ is zero, the
average of the \emph{physical} spatial curvature $\avg{\Cal{R}}$
\emph{does not} vanish, and is given by (see Eqn. \eqref{lumin7})
$\avg{\Cal{R}}=-5\beta/\aD$. If we take $\beta>0$, which guarantees
that the scale factor will accelerate for $\aD>(4\pi
G\rho_i/\beta)^{1/2}$ (see Eqn. \eqref{lumin8b}), then we have
$\avg{\Cal{R}}<0$. The quantity $\alpha(z)$ is now given by  
\begin{equation}
\alpha(z)=\int_{\aD(z)}^{\uDnow{a}}{\frac{d\aD}{\left( C_1\aD+C_2\aD^3 
    \right)^{1/2}}} ~~;~~  C_1=\frac{8\pi G}{3}\rho_i     ~,~
    C_2=\frac{2\beta}{3} \,. 
\label{lumin16-1}
\end{equation}
Since $C_1$ and $C_2$ are positive, this integral is well behaved for
all $z$. Further, since $\uD{k}=0$, we have
$\dlz=\uDnow{a}(1+z)\alpha(z)$. Defining
$\Omega_{\Cal{Q}0}\equiv(2/3)(\uDnow{\Cal{Q}}/\uDnow{H}^2)$ and noting 
that $C_1=\Omega_{m0}\uDnow{H}^2\uDnow{a}^3$ and
$C_2=(2/3)\uDnow{\Cal{Q}}\uDnow{a} =
\Omega_{\Cal{Q}0}\uDnow{H}^2\uDnow{a}$, we get  
\begin{equation}
\dlz=\frac{(1+z)}{\uDnow{H}}\int_0^z{\frac{d\tilde z}{\left[
      \Omega_{m0}(1+\tilde z)^3 + \Omega_{\Cal{Q}0}(1+\tilde z)
      \right]^{1/2}}}   \,.
\label{lumin16-2}
\end{equation}
In this notation, which is reminiscent of that in the standard LCDM
model, if we use Eqn. \eqref{lumin8a} evaluated at present time
we also see that $\Omega_{m0}+\Omega_{\Cal{Q}0}=1$, and that
acceleration begins at a redshift given by 
\begin{equation}
1+z_{\rm acc} = \uDnow{a}\left(\frac{\beta}{4\pi G\rho_i}\right)^{1/2}
= \left(\frac{\beta/\uDnow{a}}{4\pi G\rho_0}\right)^{1/2} 
=\left(\frac{\Omega_{\Cal{Q}0}}{\Omega_{m0}}\right)^{1/2} \,.
\label{lumin16-3}
\end{equation}
\begin{figure}[t]
\centering
\includegraphics[angle=-90,width=.75\textwidth]{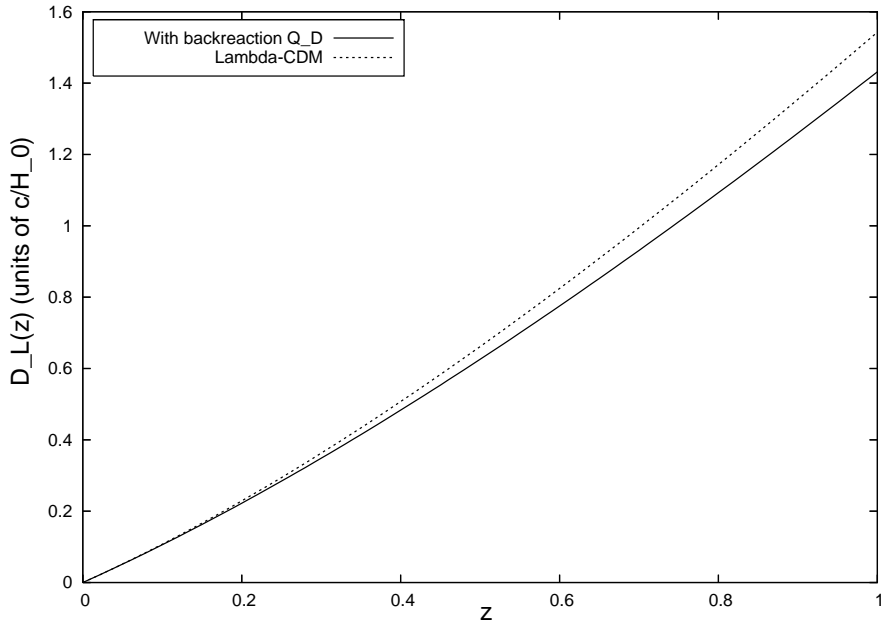}
\caption{The luminosity distance (in units of $c\uDnow{H}^{-1}$)
  versus redshift for the backreaction model with $n=-1$ defined in
  Eqns. \eqref{lumin16-1} and \eqref{lumin16-2} (solid line) and for
  the standard LCDM model with $\Omega_m=0.3$ and $\Omega_\Lambda=0.7$
  (dotted line).}    
\label{fig2}
\end{figure}
As an example, in fig. \ref{fig2} we have plotted the behaviour of
$\dlz$ in units of $c\uDnow{H}^{-1}$ in this model (for redshifts
between $z=0$ and $z=1$), setting $\Omega_{m0}=0.3$,
$\Omega_{\Cal{Q}0}=0.7$. In this case the acceleration redshift is
$z_{\rm acc}\simeq0.53$. For comparison we also show the \dlz\ curve
in the standard LCDM model with $\Omega_m=0.3$ and
$\Omega_\Lambda=0.7$, for which the acceleration redshift is $z_{\rm
  acc}=(2\Omega_\Lambda/\Omega_m)^{1/3}-1\simeq0.67$. [Interestingly, a
model in which the backreaction $\uD{\Cal{Q}}$ behaves as $\aD^{-1}$
at early times, has recently been studied by Li and Schwarz
\cite{li-schwarz} in the context of spatial averaging of a perturbed
FLRW Cosmological model.] \\~\\     

We have therefore demonstrated the existence of models of the
backreaction which can admit an accelerating scale factor. It is
important to keep in mind though, that the models considered
above were not physically well-motivated, they served only to
demonstrate the construction of luminosity distances using the
modified dynamics. A more realistic scenario would probably contain a
backreaction which becomes significant only at late times, when
structure formation has advanced sufficiently far. It will be an
interesting exercise to numerically construct luminosity distances
for such models and compare them with those of the concordance model. 
\label{lumin}

\section{Discussion}
We have argued that for the purposes of Cosmological study, an
explicit coarse graining of the matter density in the Universe must be
performed using the formalism of spatial averaging developed by
Buchert. Due to the observed large scale homogeneity of the matter
density and the isotropy of the CMB radiation, the notionally
smoothed out template metric obtained after this coarse graining must
be described by a FLRW line element, with the scale factor being given by
Buchert's functional $\aD(t)$. The procedure for constructing
luminosity distances is therefore identical to that in the standard
approach, except that the \emph{evolution} of the effective scale
factor must be given by the modified FLRW equations \eqref{avg7}
supplemented by the integrability condition \eqref{avg9}. (The scale
factor and its evolution will be further modified by including perfect
fluid sources with non-zero pressure \cite{buch4} and the curvature 
backreaction and volume effects \cite{buchRG}.) Such a construction of
luminosity distances must be interpreted as a convenient way of
comparing observations with theoretical models, while retaining the
essence of the physical inhomogeneous geometry. It will be
an interesting project to compare such luminosity distances in
realistic backreaction scenarios with current Supernovae observations
in an attempt to constrain the magnitude and evolution of the
backreaction.    

We end by discussing some additional issues concerning an explicit
smoothing of the metric which we avoided in the present
treatment, and also some potential problems with our
construction. Firstly, we return to the issue of holding the comoving
range $R_{hom}$, and in particular the smoothing scale $L$, 
constant. This is justified only so long as the condition $L \ll
L_{\rm Hubble}(z)$ is satisfied, where $L_{\rm Hubble}(z)$ is the
appropriate comoving scale associated with the observable Universe at
redshift $z$. For example, if we choose a scale $L$ which satisfies
this condition at the present epoch, the same scale will not in
general be valid at the last scattering epoch of the CMB. However, it
seems reasonable that for a small enough redshift range, the condition
$L \ll L_{\rm Hubble}(z)$ can be satisfied with a single constant
$L$. For epochs sufficiently far back in the past, the scale $L$ would
have to be reduced, and the construction of the \dlz\ relation would
no longer be valid. 

Further, a change in the size of the averaging domain would in general
affect both the scale factor \aD\ and the ``constant'' \uD{k}\ in a
non-trivial way. In particular, any change in \uD{k}\ would render the
effective FLRW metric \eqref{smop1} meaningless. Also, a change in
\uD{k}\ need not necessarily be caused by a change in the averaging
domain size. It seems likely \cite{buch-corresp} that an explicit
smoothing of the geometry using Buchert and Carfora's technique will
yield a 3-space of different spatially constant curvatures at
different times, this difference being independent of any change in
the averaging domain size with time and being possibly incommensurate
with our assumption of $\Cal{R}_{\rm eff}\propto\aD^{-2}$. This could
be interpreted in terms of a change in \uD{k}\ with time, and in such
a case, one would be dealing with a \emph{different} FLRW template at 
different times. The effects of such a construction are not clear at
present. Clearly this issue needs more careful study, and we hope to
address this in the future. Of course, if \uD{k}\ is made a function
of time, care must be taken that the matter content is now so chosen that
Einstein's equations continue to be satisfied, as pointed out by
R\"{a}s\"{a}nen \cite{r1} in response to \cite{Hou}. 

Also, as pointed out by R\"{a}s\"{a}nen \cite{ras}, the use of a
homogeneous and isotropic FLRW template will necessarily ignore
physical effects such as a non-trivial shear in the underlying
geometry. A potential problem with our template construction is that
effects of the non-trivial scalar curvature $^{(3)}\Cal{R}(\vec{x},t)$
on light propagation are also simplified to a proportionality to
$\aD^{-2}$. Although the full scalar curvature will show up in the
evolution of $\aD(t)$ via its spatial average \avg{\Cal{R}}, it is
possible that this construction may inaccurately model the real
Universe. A study of the evolution of the effective spatial curvature
as mentioned in the preceding paragraph may help resolve this issue.

There is also the related issue of degeneracy, namely that several
underlying inhomogeneous models may reproduce the same backreaction
effects when averaged. For example, it was shown \cite{aptp} in the
context of the LTB solutions, that models with a vanishing spatial
curvature (the so called marginally bound models) in the inhomogeneous
geometry also have a vanishing backreaction \uD{\Cal{Q}} after
averaging. One is now faced with the situation wherein interpreting
say luminosity distance observations within the inhomogeneous geometry
may distinguish two inhomogeneous models, but interpreting the data
\emph{after} averaging would not make this distinction and would hence
perhaps rule out \emph{both} models. This issue has been recently
highlighted by Enqvist and Mattsson (the last paper in
Ref. \cite{LTB}) who show that a marginally bound LTB model can fit
the data from Type Ia SNe without any spatial averaging. Since the
standard FLRW matter dominated model is also (a special, homogeneous
case of) a marginal LTB model which \emph{doesn't} fit the SNe data
\cite{riess}, the formalism after spatial averaging would rule out
both models by demanding a non-trivial backreaction instead. However,
since the LTB (and in fact any inhomogeneous) models themselves face a
similar degeneracy issue (for example several different LTB models can
currently claim to fit the SNe data \cite{LTB}), this matter cannot be
considered as settled, and also deserves further attention.  

Lastly, we note that Buchert's averaging scheme itself has been
criticized in the literature (see, e.g., Ref. \cite{wald}) on the
grounds that the effects of averaging in a noncovariant manner as done
in this scheme, are likely to be gauge artifacts. 
It should be noted that within second-order perturbation theory,
gauge invariance of the averaged equations has been demonstrated in
\cite{kolb} and \cite{li-schwarz}.  
It is also possible to
approach the problem of determining modifications to the standard
Cosmological equations via a covariant method such as Zalaletdinov's
Macroscopic Gravity \cite{zala} for example. Such an attempt has in
fact been made recently and has led to effective Cosmological
equations very similar to those of Buchert (see
Ref. \cite{spatlim}). The proposal in the present paper however, is
attractive to us due to its simplicity in implementation, and may be
treated as a phenomenological model of the more rigorous equations
derived in Ref. \cite{spatlim}.
\label{discuss}

\section*{Acknowledgments}
We have greatly benefited from our correspondence with Thomas Buchert
and Syksy R\"{a}s\"{a}nen, and it is a pleasure to acknowledge their
insightful comments and criticism of an earlier draft. AP also thanks
Rakesh Tibrewala for useful and stimulating discussions.

\label{refer}

\end{document}